\documentclass[aps,prl,reprint,showpacs,amsfonts,amssymb,superscriptaddress,floatfix,twocolumn]{revtex4-1}
\usepackage{graphicx}
\usepackage{amsmath}
\DeclareMathOperator\arctanh{arctanh}
\usepackage{dcolumn}
\usepackage{color}
\usepackage{bm}
\usepackage{braket}
\usepackage{float}
\usepackage{calc}
\usepackage{MnSymbol}
\newcommand{\shiftleft}[2]{\makebox[0pt][r]{\makebox[#1][l]{#2}}}

\begin{document}

\title{Instability of Rotationally Tuned Dipolar Bose-Einstein Condensates}

\author{S. B. Prasad}
\affiliation{School of Physics, University of Melbourne, Melbourne, 3010, Australia}
\author{T. Bland}
\affiliation{Joint Quantum Centre Durham-Newcastle, School of Mathematics, Statistics and Physics, Newcastle University, Newcastle upon Tyne, NE1 7RU, United Kingdom}
\author{B. C. Mulkerin}
\affiliation{Centre for Quantum and Optical Science, Swinburne University of Technology, Melbourne, 3122, Australia}
\author{N. G. Parker}
\affiliation{School of Physics, University of Melbourne, Melbourne, 3010, Australia}
\affiliation{Joint Quantum Centre Durham-Newcastle, School of Mathematics, Statistics and Physics, Newcastle University, Newcastle upon Tyne, NE1 7RU, United Kingdom}
\author{A. M. Martin}
\affiliation{School of Physics, University of Melbourne, Melbourne, 3010, Australia}

\date{\today}

\begin{abstract}
The possibility of effectively inverting the sign of the dipole-dipole interaction, by fast rotation of the dipole polarization, is examined within a harmonically trapped dipolar Bose-Einstein condensate. Our analysis is based on the stationary states in the Thomas-Fermi limit, in the corotating frame, as well as direct numerical simulations in the Thomas-Fermi regime, explicitly accounting for the rotating polarization. The condensate is found to be inherently unstable due to the dynamical instability of collective modes. This ultimately prevents the realization of robust and long-lived rotationally tuned states. Our findings have major implications for experimentally accessing this regime.
\end{abstract}

\maketitle
Dipolar Bose-Einstein condensates (BECs) have proved to be a unique, highly-controllable platform for studying the interplay of quantum many-body physics and magnetic interactions~\cite{prl_94_16_160401_2005, pra_77_6_061601r_2008, prl_106_1_015301_2011, prl_107_19_190401_2011, prl_108_21_210401_2012, nature_530_7589_194-197_2016}. Compared to conventional condensates, in which the atoms undergo short-range, contact-like, isotropic inter-particle interactions, a dipolar BEC also enjoys a long-range anisotropic dipole-dipole interaction (DDI)~\cite{physrep_464_3_71-111_2008, repprogphys_72_12_126401_2009, njp_11_5_055049_2009, chemrev_112_9_5012-5061_2012, jphyscondesmatter_29_10_103004_2017}. While the first dipolar BECs were realised in a gas of $^{52}$Cr~\cite{prl_94_16_160401_2005,pra_77_6_061601r_2008}, atoms with larger magnetic dipole moments such as $^{164}$Dy~\cite{prl_107_19_190401_2011,nature_530_7589_194-197_2016} and $^{168}$Er~\cite{prl_108_21_210401_2012} have now been cooled to form strongly dipolar BECs. This has led to the observation of magnetostriction~\cite{nature_448_7154_672-675_2007}, dipole mediated-stability~\cite{prl_101_8_080401_2008,natphys_4_3_218-222_2008}, anisotropic superfluidity~\cite{prl_121_3_030401_2018}, dipolar gap solitons in 2D free space~\cite{pra_95_6_063613_2017}, the roton mode~\cite{prl_90_25_250403_2003, prl_98_3_030406_2007, natphys_14_5_442-446_2018} and the discovery of self-bound dipolar droplets~\cite{nature_530_7589_194-197_2016, nature_539_7628_259-262_2016, prl_116_21_215301_2016, prx_6_4_041039_2016}.

The net interaction potential, combining the contact interactions and DDI, is pivotal to both the stability of the gas and to realize regimes of novel physics. To date, experimentalists have controlled this potential by exploiting Feshbach resonances to tune the contact interaction, thereby allowing for DDI-dominated regimes~\cite{prl_101_8_080401_2008, natphys_4_3_218-222_2008}. Furthermore it has been suggested that the magnitude and sign of the DDI can be tuned via rotation of the dipole moments \cite{prl_89_13_130401_2002}. Consider a BEC of bosons with magnetic dipole moment $\mu_\text{d}$, polarized uniformly along an axis $\hat{\mathbf{e}}(t)$, such that the DDI is
\begin{equation}
U_{\text{dd}}\left(\mathbf{r},t\right) = \frac{C_{\text{dd}}}{4\pi}\frac{1-3\left(\hat{\mathbf{e}}(t)\cdot\mathbf{r}\right)^2}{|\mathbf{r}|^3}, \label{eq:ddint}
\end{equation}
where $C_{\text{dd}} = \mu_0\mu_\text{d}^2$ and $\mu_0$ is the vacuum permeability. If $\hat{\mathbf{e}}(t)$ rotates about the $z$-axis at a tilt angle $\varphi$, the time averaged DDI over one rotation cycle is \cite{prl_89_13_130401_2002}
\begin{equation}
\llangle U_{\text{dd}}\left(\mathbf{r}\right)\rrangle = \frac{C_{\text{dd}}}{4\pi}\left(\frac{3\cos^2\varphi - 1}{2}\right)\left[\frac{1-3\left(\hat{z}\cdot\mathbf{r}\right)^2}{|\mathbf{r}|^3}\right]. \label{eq:ddtimeavg}
\end{equation}
Thus, in the rapid-rotation limit, the tilt angle $\varphi$ may be used to tune the effective strength of the DDI and in particular, when $\cos^2\varphi > 1/3$, the effective DDI strength becomes negative, corresponding to an unusual `anti-dipolar' regime in which side-by-side alignment of the dipole moments is energetically preferred to head-to-tail alignments. Subsequent theoretical studies of dipolar BECs in this regime, which invoked the rotational tuning mechanism by setting $C_{\text{dd}} < 0$, led to predictions of novel physics such as molecular bound states in dark solitons \cite{bland2015controllable}, multi-dimensional dark \cite{nath2008stability} and bright \cite{pedri2005two, njp_19_2_023019_2017} solitons, stratified turbulence \cite{prl_121_17_174501_2018} and the roton instability of vortex lines~\cite{prl_100_24_240403_2008}. In this direction, a recent experimental study of rotational tuning by Tang {\it et al.}~\cite{prl_120_23_230401_2018} has reported a realization of the anti-dipolar regime.

In this work we revisit rotational tuning of a dipolar BEC, in a cylindrically symmetric harmonic trap of the form $V_{\text{T}}\left(\mathbf{r}\right) = \frac{1}{2}m\omega_{\perp}^2\left(x^2 + y^2 + \gamma^2 z^2\right)$, and model polarization rotational angular frequencies, $\Omega$, greater than $\omega_{\perp}$. This rotation is seen to result in an asymmetry of the condensate about $\hat{z}$ that is not evident if the DDI due to a rotating polarization is directly replaced by its time-averaged counterpart. This asymmetry results in a dynamical instability, similar to those predicted~\cite{prl_86_3_377-390_2001, prl_87_19_190402_2001, pra_73_6_061603r_2006} and observed~\cite{prl_84_5_806-809_2000, prl_88_1_010405_2001} for non-dipolar condensates in rotating ellipsoidal traps, that prevents the formation of a dynamically-stable rotationally-tuned state. This is elucidated via two distinct approaches, one being based on a semi-analytical treatment in the \textit{Thomas-Fermi} (TF) limit, and the other being time-dependent numerical simulations. These results raise major questions over the pursuit of rotational tuning of dipolar BECs.

As the maximally anti-dipolar regime occurs for $\varphi = \pi/2$, we consider the polarizing field $\hat{\mathbf{e}}$ to be rotating in the $x-y$ plane at angular frequency $\Omega$, and work in a reference frame co-rotating with $\mathbf{e}$ such that we may fix $\hat{\mathbf{e}} = \hat{x}$ in this frame. Then, the dipolar condensate order parameter $\psi(\mathbf{r},t)$~\cite{prl_85_9_1791-1794_2000} for bosons of mass $m$ can be modeled by the dipolar Gross-Pitaevskii Equation (dGPE),
\begin{equation}
i\hbar\frac{\partial\psi}{\partial t} = \left[-\frac{\hbar^2\nabla^2}{2m} + V_{\text{T}} + V_{\text{int}} + i\hbar\Omega\left(x\frac{\partial}{\partial y}-y\frac{\partial}{\partial x}\right)\right]\psi.  \label{eq:gpe}
\end{equation}
Assuming that the rotation frequency is slow enough that the dipole moments remain aligned along $\hat{\mathbf{e}}$ at all times, the interaction potential $V_{\text{int}}$ is specified by~\cite{pra_63_5_053607_2001}
\begin{equation}
V_{\text{int}}(\mathbf{r},t) = g\left|\psi(\mathbf{r},t)\right|^2 + \int\mathrm{d}\mathbf{r'}\,U_{\text{dd}}(\mathbf{r}-\mathbf{r}')\left|\psi(\mathbf{r'},t)\right|^2, \label{eq:ddpot}
\end{equation}
where $g = 4\pi\hbar^2a_{\text{s}}/m$ and $a_{\text{s}}$ is the bosonic s-wave scattering length. The DDI strength may be related to $a_{\text{s}}$ through a dimensionless ratio $\epsilon_{\text{dd}} = C_{\text{dd}}/(3g)$.

Previous studies of rotationally-tuned dipolar BECs have involved setting $\Omega = 0$ and replacing $U_{\text{dd}}$ with $\llangle U_{\text{dd}}\rrangle$ in Eq.~\eqref{eq:ddpot}. To test the validity of this procedure, we solve for the stationary solutions of Eq.~\eqref{eq:gpe}, which obey $i\hbar\partial_t\psi = \mu\psi$, with $\mu$ being the condensate's chemical potential. We re-express the order parameter as $\psi = \sqrt{n}\exp(iS)$, where $S$ is the condensate phase and $n$, the condensate density, is normalized to the condensate number $N$ via $\int\mathrm{d}^3r\,n(\mathbf{r}) = N$. In the TF limit, obtained by neglecting the zero-point kinetic energy of the condensate~\cite{pra_78_4_041601r_2008}, the stationary solutions are of the form
\begin{gather}
n_{\text{TF}}(\mathbf{r}) = n_0\left(1-\frac{x^2}{\kappa_x^2R_z^2}-\frac{y^2}{\kappa_y^2R_z^2}-\frac{z^2}{R_z^2}\right). \label{eq:tfdensity} \\
S_{\text{TF}}\left(\mathbf{r},t\right) = \alpha xy - \mu t/\hbar, \label{eq:tfphase}
\end{gather}
Here $n_0 = 15N/(8\pi\kappa_x\kappa_yR_z^3)$ is the peak density, $R_i$ is the TF radius of the dipolar BEC along the $i$-axis, and $\kappa_x = R_x/R_z$ and $\kappa_y = R_y/R_z$ are the condensate aspect ratios, with respect to $\hat{z}$, along $\hat{x}$ and $\hat{y}$ respectively.

\begin{figure}
\includegraphics[width=\linewidth]{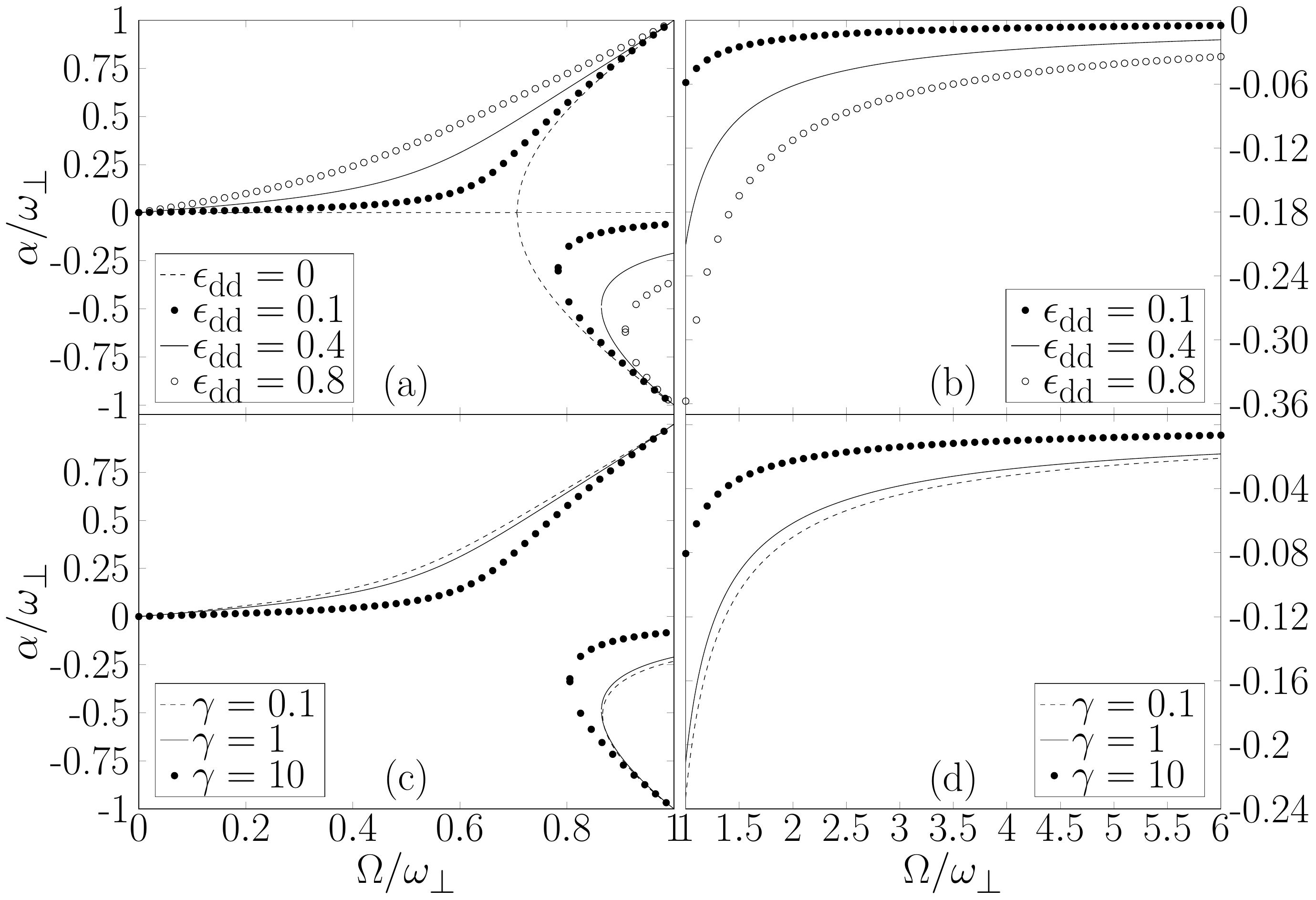}
\vspace*{-5mm}
\caption{Stationary solutions, as characterized by $\alpha$, as a function of $\Omega$: (a, b) $\gamma = 1$ and various $\epsilon_{\text{dd}}$; (c, d) $\epsilon_{\text{dd}} = 0.4$ and various $\gamma$. In (b), the $\gamma = 1$, $\epsilon_{\text{dd}} = 0$ branch has $\alpha = 0$.}
\label{bothrangeomega0to6profile}
\end{figure}

The TF stationary solutions are uniquely determined by a set of consistency relations, whose derivation is presented in the Supplemental Material. For a given choice of $\lbrace\gamma, \epsilon_{\text{dd}}, \Omega, \omega_{\perp}\rbrace$, the consistency relations are given by:
\begin{align}
\kappa_x^2 &= \frac{1}{\zeta}\left(\frac{\omega_{\perp}\gamma}{\widetilde{\omega}_x}\right)^2\left[1 + \epsilon_{\text{dd}}\left(\frac{9}{2}\kappa_x^3\kappa_y\beta_{200}-1\right)\right], \label{eq:kappax} \\
\kappa_y^2 &= \frac{1}{\zeta}\left(\frac{\omega_{\perp}\gamma}{\widetilde{\omega}_y}\right)^2\left[1 + \epsilon_{\text{dd}}\left(\frac{3}{2}\kappa_y^3\kappa_x\beta_{110}-1\right)\right], \label{eq:kappay} \\
0 &= (\alpha + \Omega)\left[\widetilde{\omega}_x^2 - \frac{9}{2}\epsilon_{\text{dd}}\frac{\omega_{\perp}^2\kappa_x\kappa_y\gamma^2}{\zeta}\beta_{200}\right] \nonumber \\
&+ (\alpha - \Omega)\left[\widetilde{\omega}_y^2 - \frac{3}{2}\epsilon_{\text{dd}}\frac{\omega_{\perp}^2\kappa_x\kappa_y\gamma^2}{\zeta}\beta_{110}\right]. \label{eq:alphakappa}
\end{align}
Here, $\alpha, \widetilde{\omega}_x, \widetilde{\omega}_y, \beta_{ijk}$ and $\zeta$ are defined by the following:
\begin{gather}
\alpha = \frac{\kappa_x^2 - \kappa_y^2}{\kappa_x^2 + \kappa_y^2}\Omega, \label{eq:alphadefn} \\
\widetilde{\omega}_x^2 = \omega_{\perp}^2 + \alpha^2 - 2\alpha\Omega\,,\,\widetilde{\omega}_y^2 = \omega_{\perp}^2 + \alpha^2 + 2\alpha\Omega, \label{eq:omegay} \\
\beta_{ijk} = \int_0^{\infty}(s+\kappa_x^2)^{-i-\frac{1}{2}}(s+\kappa_y^2)^{-j-\frac{1}{2}}(s+1)^{-k-\frac{1}{2}}\,\mathrm{d}s, \label{eq:beta} \\
\zeta = 1 + \epsilon_{\text{dd}}\left(\frac{3}{2}\kappa_x\kappa_y\beta_{101} - 1\right).
\end{gather}
Equation~\eqref{eq:alphadefn} encapsulates the in-plane anisotropy of the stationary TF density as a function of $\Omega$, with a positive (negative) $\alpha$ implying that the condensate density is elongated along $\hat{x}$ ($\hat{y}$). Figure~\ref{bothrangeomega0to6profile} shows how $\alpha$ varies with $\Omega$ for (a, b) various values of $\epsilon_{\text{dd}}$ while fixing $\gamma$, and for (c, d) various values of $\gamma$ while fixing $\epsilon_{\text{dd}}$. If $\epsilon_{\text{dd}} = 0$, $\alpha = 0$ is a valid solution for all $\Omega$. A bifurcation occurs at $\Omega = \Omega_b$, with the addition of two new branches, symmetric about the $\Omega$-axis, that exist only for $\Omega_b \leq \Omega < \omega_{\perp}$. The symmetry about the $\Omega$-axis is broken when $\epsilon_{\text{dd}} > 0$. Instead, we have $\alpha > 0$ when $0 < \Omega < \Omega_b$, and this branch persists for $\Omega_b \leq \Omega < \omega_{\perp}$. The bifurcation is now in the form of two additional $\alpha < 0$ solutions for $\Omega \geq \Omega_b$, which are simply connected to each other at $\Omega = \Omega_b$. The two branches with the highest $|\alpha(\Omega)|$ terminate at $\Omega = \omega_{\perp}$. This is characteristic of the TF limit in a rotating frame, with similar bifurcations occurring in a BEC rotating about the $z$-axis with a planar trapping ellipticity, with or without $z$-polarised dipoles~\cite{prl_86_3_377-390_2001, prl_86_20_4443-4446_2001, prl_98_15_150401_2007, pra_80_3_033617_2009}. This has been attributed to the $L_z = 2$ quadrupole mode being energetically unstable for $\Omega > \Omega_b$, resulting in $\Omega_b(\epsilon_{\text{dd}} = 0) = \omega_{\perp}/\sqrt{2}$ as demonstrated in Fig.~\ref{bothrangeomega0to6profile} (a) \cite{prl_86_3_377-390_2001,pitaevskiistringaribec}.

It is evident that these rotating-frame solutions tend towards cylindrical symmetry ($\alpha = 0$) as $\Omega\rightarrow\infty$. We proceed to test whether they agree with the non-rotating TF stationary solutions found by utilising the time-averaged DDI. The latter are exactly symmetric about $\hat{z}$ and possess an aspect ratio $\kappa_{\parallel} \equiv \kappa_x = \kappa_y$ specified via~\cite{prl_92_25_250401_2004, pra_71_3_033618_2005}
\begin{gather}
3\epsilon_{\text{dd}}\kappa_{\parallel}^2\left[\left(1+\frac{\gamma^2}{2}\right)\frac{f(\kappa_{\parallel})}{1-\kappa^2} - 1\right] = (\epsilon_{\text{dd}} + 2)(\gamma^2 - \kappa_{\parallel}^2), \label{eq:pfausol} \\
f(\kappa_{\parallel}) = \frac{1+2\kappa_{\parallel}^2}{1-\kappa_{\parallel}^2} - \frac{3\kappa_{\parallel}^2\arctanh\sqrt{1-\kappa_{\parallel}^2}}{(1-\kappa_{\parallel}^2)^{3/2}}. \label{eq:dipolefunc}
\end{gather}
We compare this with the true time-averaged condensate density by transforming Eq. \eqref{eq:tfdensity} to the laboratory coordinates and time-averaging over one rotation cycle, yielding the time-averaged aspect ratio $\kappa_{\perp} = \sqrt{2}(\kappa_x^{-2} + \kappa_y^{-2})^{-1/2}$.

\begin{figure}
\includegraphics[width=\linewidth]{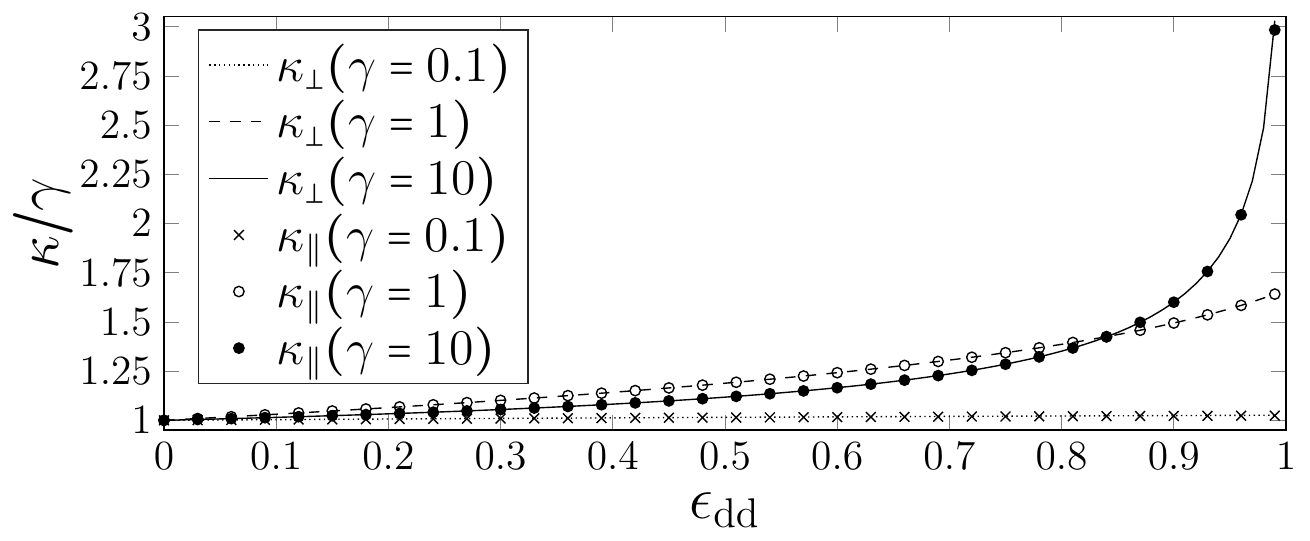}
\vspace*{-5mm}
\caption{Comparison of the solutions for the rapid-rotating dipoles and time-averaged DDI formalisms. Shown is $\kappa_{\perp}$ (lines) evaluated at $\Omega = 50\omega_{\perp}$ and $\kappa_{\parallel}$ (markers), with $\epsilon_{\text{dd}}\rightarrow -\epsilon_{\text{dd}}/2$, as a function of $\epsilon_{\text{dd}}$ and for various trap aspect ratios. }
\label{gammarangeomega50edd0to1timeavg}
\end{figure}

Figure~\ref{gammarangeomega50edd0to1timeavg} compares $\kappa_{\perp}$ and $\kappa_{\parallel}$ as a function of $\epsilon_{\text{dd}}$, with a range of trapping aspect ratios, $\gamma$, being considered and with $\kappa_{\perp}$ evaluated at a suitably high rotation frequency ($\Omega = 50\omega_{\perp}$). An almost perfect agreement between the two methodologies is evident. Note that when $\gamma = 1$, the condensate is flattened with respect to the $z$-axis, consistent with an effective side-by-side orientation of $z$-polarized dipole moments.

To verify the stationary solutions, Eqs.~\eqref{eq:kappax} -- \eqref{eq:alphakappa}, we numerically solve the $3$D dGPE for a dipolar BEC of $N = 10^5$ bosons; $N$ is chosen to be sufficiently large for a meaningful comparison with the TF analysis~\cite{pra_78_4_041601r_2008}. With $\Omega$ and $\gamma$ fixed throughout, our initial condition is the stationary state for $\epsilon_{\text{dd}} = 0$ obtained by imaginary time propagation of the dGPE~\cite{kinetrelatmod_6_1_1-135_2013}. In time, $\epsilon_{\text{dd}}$ is slowly ramped up (at a rate $\mathrm{d}\epsilon_{\text{dd}}/\mathrm{d}t = 10^{-3}\omega_{\perp}$), such that the condensate can slowly traverse the corresponding stationary solutions to high adiabaticity. Further details regarding the simulation are provided in the Supplemental Material.

\begin{figure}
\includegraphics[width=\linewidth]{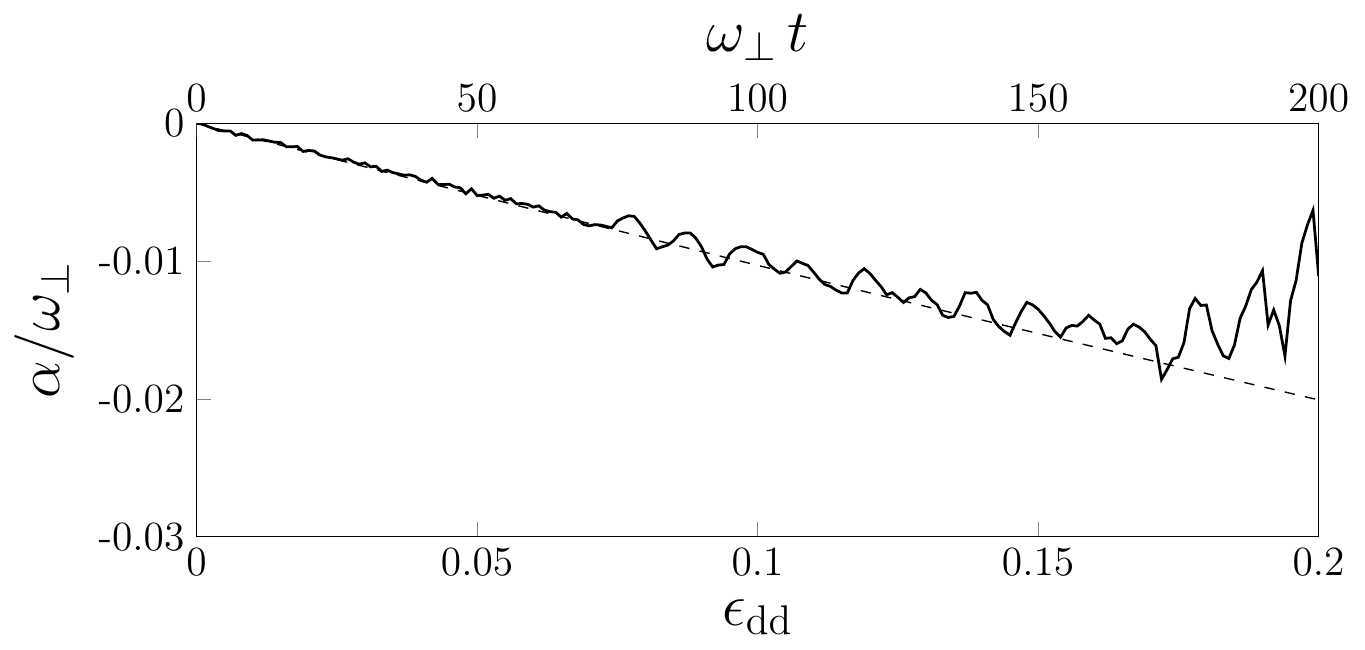}
\vspace*{-5mm}
\caption{Comparison between the numerical simulation and the TF solutions: $\alpha$ as a function of $\epsilon_{\text{dd}}$ at $\Omega = 3\omega_{\perp}$ and $\gamma = 1$, determined via Eqs.~\eqref{eq:kappax} -- \eqref{eq:alphakappa} (dashed line) and numerical simulation for $N = 10^5$ (solid line).}
\label{gpevstf}
\end{figure}

Figure~\ref{simulationimages} depicts the density and phase, during a simulation with fixed $\Omega = 3\omega_{\perp}$ and $\gamma = 1$, as cross-sections at $z = 0$ taken at times when $\epsilon_{\text{dd}} = 0.05$, $0.15$, and $0.20$. For low $\epsilon_{\text{dd}}$, the condensate is consistent with the TF stationary solution: the density is smooth and approximates the paraboloid profile of Eq.~\eqref{eq:tfdensity}, while the phase approximates the quadrupolar flow of Eq.~\eqref{eq:tfphase}. However, for higher $\epsilon_{\text{dd}}$, the density and phase profiles deviate considerably from this form, first visible through a rippling of the density being evident ($\epsilon_{\text{dd}} = 0.15$), which later evolves towards a fragmented state ($\epsilon_{\text{dd}} = 0.2$). Figure~\ref{gpevstf} tracks this departure from the TF solution by comparing $\alpha$ as determined from the simulation with that found from Eqs.~\eqref{eq:kappax} -- \eqref{eq:alphakappa}. While the agreement is excellent at low $\epsilon_{\text{dd}}$, i.e. early time, the numerical value begins to fluctuate at $\epsilon_{\text{dd}}\approx 0.075$. The amplitude of this fluctuation grows with time, with the numerical solution diverging from the semi-analytical one entirely when $\epsilon_{\text{dd}}\approx 0.17$.

\begin{figure}
\begin{center}
  \begin{tabular}{c c c c c c c}
    & & $\epsilon_{\text{dd}}=0.05$ & $\epsilon_{\text{dd}}=0.15$ & $\epsilon_{\text{dd}}=0.2$ & & \\ \cline{3-5}
    \rotatebox[origin=c]{90}{$n(x,y,z=0)\,/\,n_0$}
    &
    \rotatebox[origin=c]{90}{$0$ \hspace{1.1cm} $0.003$}
    \begin{minipage}[c][0.20\columnwidth][c]{0.05\columnwidth}
    	\centering
    	\includegraphics[width=0.4\columnwidth]{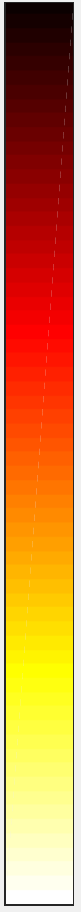}
    \end{minipage} 
    &
    \begin{minipage}[c][0.20\columnwidth][c]{0.20\columnwidth}
    \centering
      \includegraphics[trim={0.1cm 0.1cm 0.1cm 0.15cm},clip,width=0.8\columnwidth]{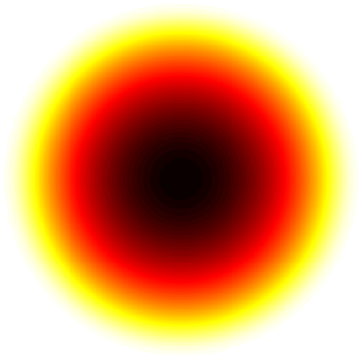}
    \end{minipage}
    &
    \begin{minipage}[c][0.2\columnwidth][c]{0.2\columnwidth}
    \centering
    \includegraphics[trim={0.5cm 0.6cm 0.5cm 0.5cm},clip,width=0.8\columnwidth]{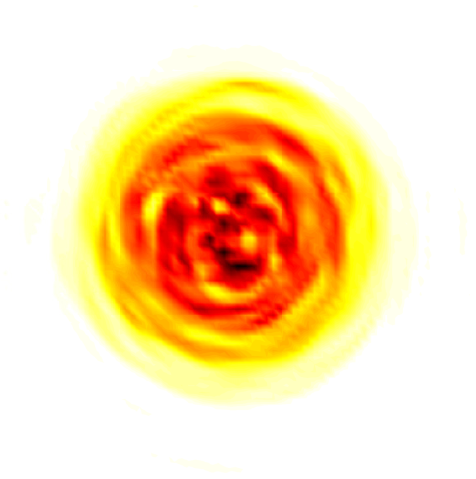}
    \end{minipage}
    & 
    \begin{minipage}[c][0.2\columnwidth][c]{0.2\columnwidth}
    \centering
	\includegraphics[trim={2.7cm 2.7cm 2.7cm 2.3cm},clip,width=0.80\columnwidth]{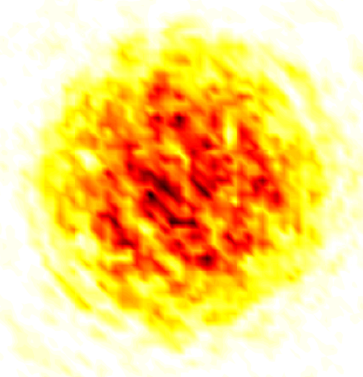}
    \end{minipage}
        &
    \rotatebox[origin=c]{90}{$-8$ \hspace{0.5cm}0\hspace{0.5cm} $8$\color{white}{$-$}}
    & \rotatebox[origin=c]{90}{$y\,/\,l_\perp$}
    \\ \cline{3-5}
        \rotatebox[origin=c]{90}{$S(x,y,z=0)$}
        &
        \rotatebox[origin=c]{90}{$0$ \hspace{1.3cm} $2\pi$}
    \begin{minipage}[c][0.20\columnwidth][c]{0.05\columnwidth}
    	\centering
    	\includegraphics[width=0.4\columnwidth]{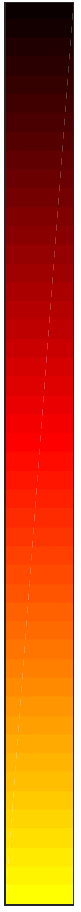}
    \end{minipage} 
    &
    \begin{minipage}[c][0.2\columnwidth][c]{0.2\columnwidth}
    \centering
      \includegraphics[width=0.8\columnwidth]{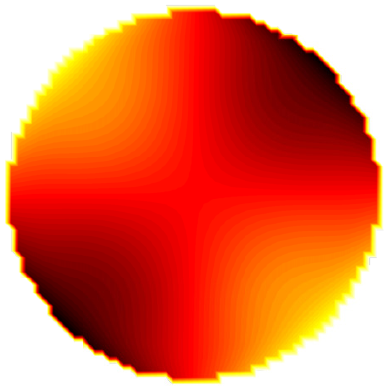}
    \end{minipage}
    &
    \begin{minipage}[c][0.2\columnwidth][c]{0.2\columnwidth}
    \centering
    \raisebox{-0.2mm}{\includegraphics[width=0.80\columnwidth]{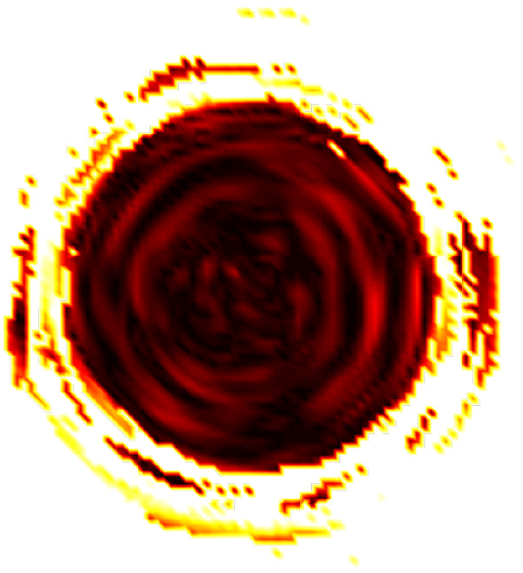}}
    \end{minipage}
    & 
    \begin{minipage}[c][0.2\columnwidth][c]{0.2\columnwidth}
    \centering
	\includegraphics[trim={0.1cm 0.1cm 0.1cm 0.1cm},clip,width=0.8\columnwidth]{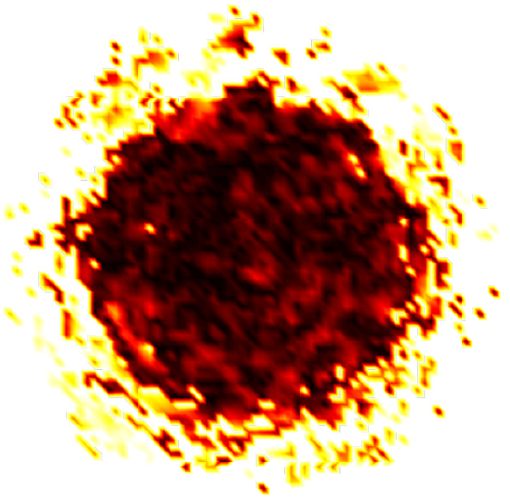}
    \end{minipage}
        &
    \rotatebox[origin=c]{90}{$-8$ \hspace{0.5cm}0\hspace{0.5cm} $8$\color{white}{$-$}}
    & \rotatebox[origin=c]{90}{$y\,/\,l_\perp$}
    \\ 
     & & $-8$ \hspace{0.5cm}0\hspace{0.5cm} $8$\color{white}{$-$} & $-8$ \hspace{0.5cm}0\hspace{0.5cm} $8$\color{white}{$-$} & $-8$ \hspace{0.5cm}0\hspace{0.5cm} $8$\color{white}{$-$} & & \\ 
     & & $x\,/\,l_\perp$ & $x\,/\,l_\perp$ & $x\,/\,l_\perp$ & &
  \end{tabular}
\end{center}
\vspace*{-0.5cm}
\caption{Simulation of a dipolar BEC during an adiabatic ramp-up of $\epsilon_{\text{dd}}$, with $N = 10^5$, $\gamma = 1$ and $\Omega = 3\omega_{\perp}$: Cross-sections at $z = 0$ of density (first row) and phase (second row) at $t = 50\omega_{\perp}^{-1}$ (first column), $t = 150\omega_{\perp}^{-1}$ (second column) and $t = 200\omega_{\perp}^{-1}$ (third column), scaled by $l_{\perp} = \sqrt{\hbar/(m\omega_{\perp})}$.  The density is normalized to $n_0=N/l_\perp^3$.}
\label{simulationimages}
\end{figure}

The deviation of the numerically-determined $\alpha$ from the semi-analytical prediction hints at the unstable growth of collective modes of the condensate~\cite{prl_105_4_040404_2010}. This motivates us to return to the TF solutions to study their response to perturbations by means of linearised perturbation analysis~\cite{prl_77_12_2360-2363_1996, prl_87_19_190402_2001, pra_80_3_033617_2009}. We proceed by writing the time-dependent density and phase as fluctuations about the respective stationary state values:
\begin{align}
n\left(\mathbf{r},t\right) &= n_{\text{TF}}(\mathbf{r}) + \delta n\left(\mathbf{r},t\right); \label{eq:nperturb} \\
S\left(\mathbf{r},t\right) &= S_{\text{TF}}\left(\mathbf{r},t\right) + \delta S\left(\mathbf{r},t\right). \label{eq:sperturb}
\end{align}
These are substituted into Eq.~\eqref{eq:gpe}, with terms quadratic (or higher) in the fluctuations being discarded. This results in an eigenvalue problem of the form~\cite{prl_87_19_190402_2001, pra_73_6_061603r_2006, prl_98_15_150401_2007, pra_80_3_033617_2009}
\begin{equation}
\frac{\partial}{\partial t}
\begin{pmatrix}
\delta S \\
\delta n
\end{pmatrix}
=
\mathcal{L}
\begin{pmatrix}
\delta S \\
\delta n
\end{pmatrix}, \label{eq:perteqns}
\end{equation}
where the explicit expression for the operator $\mathcal{L}$ is specified in the Supplemental Material. The solutions of Eq.~\eqref{eq:perteqns} are collective oscillations, and their respective eigenvalues obey
\begin{equation}
\begin{pmatrix}
\delta S\left(\mathbf{r},t\right) \\
\delta n\left(\mathbf{r},t\right)
\end{pmatrix}
\equiv
e^{\lambda t}\begin{pmatrix}
\delta S(\mathbf{r}) \\
\delta n(\mathbf{r})
\end{pmatrix}\,:\,\mathcal{L}\begin{pmatrix}
\delta S(\mathbf{r}) \\
\delta n(\mathbf{r})
\end{pmatrix} = \lambda\begin{pmatrix}
\delta S(\mathbf{r}) \\
\delta n(\mathbf{r})
\end{pmatrix}. \label{eq:eigenproblem}
\end{equation}
Examining the spectra of Eq. \eqref{eq:eigenproblem} allows for a qualitative understanding of the stability of the condensate with respect to collective modes. If a mode, indexed by $i$, features $\text{Re}(\lambda_i) > 0$, its amplitude grows exponentially and ultimately overwhelms the TF stationary solution. Thus, a stationary state is dynamically stable only if the real components of its entire spectrum is negative or zero.

Equation~\eqref{eq:eigenproblem} may be diagonalized numerically over $\mathbb{R}^3$ by utilising a polynomial basis $\lbrace x^py^qz^r\rbrace$ for $\delta S(\mathbf{r})$ and $\delta n(\mathbf{r})$~\cite{prl_87_19_190402_2001, pra_80_3_033617_2009}. Fluctuations of order $p + q + r = 0, 1, 2, \dots$ represent monopolar, dipolar and quadrupolar modes respectively, and so on~\cite{pra_82_3_033612_2010}; a rich variety of collective modes, including breathing and scissors modes, have been observed in several dipolar BEC experiments~\cite{prl_105_4_040404_2010}. Due to the inability of a numerical diagonalization scheme to explore the infinite dimensional space of polynomials, it is necessary to impose a truncation of the form $p + q + r \leq N_{\text{max}}$. However, we note that reducing the degree of the Hilbert space truncation, i.e. increasing $N_{\text{max}}$, does not modify any eigenvalues corresponding to modes with an order less than $N_{\text{max}}$, but merely increases the dimension of the truncated space of modes available to us. If, at a given point in parameter space, the diagonalization of Eq. \eqref{eq:eigenproblem} with respect to fluctuations of order less than $N_{\text{max}}$ yields at least one eigenvalue with a positive real component, it is thus sufficient to claim that the condensate is dynamically unstable up to linear order in the fluctuations.

\begin{figure}
\settototalheight{\dimen0}{\includegraphics[width=\columnwidth]{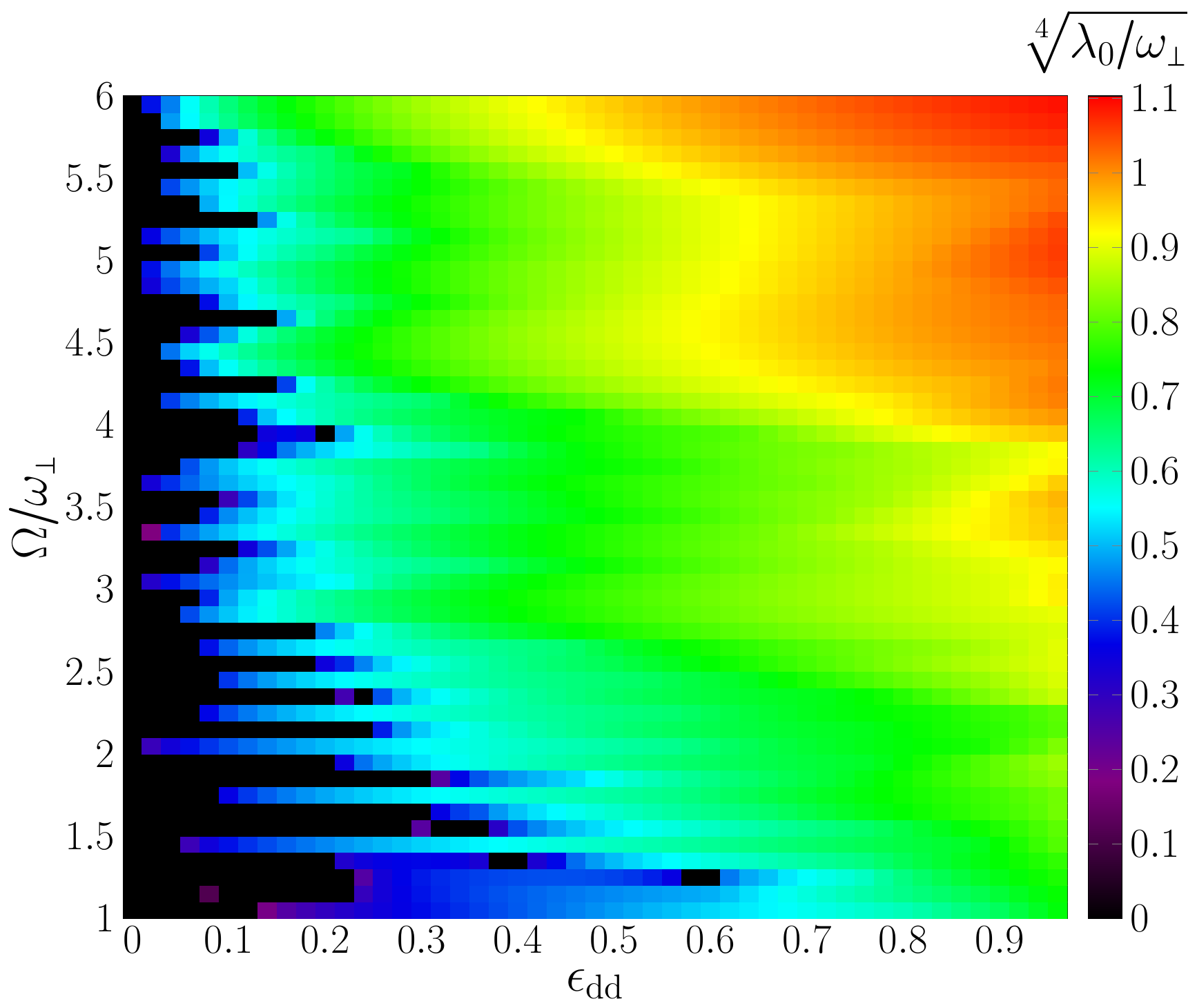}}%
\includegraphics[width=\columnwidth]{maxeiggamma1heatmap.pdf}\llap{\shiftleft{0.735\dimen0}{\raisebox{0.460\dimen0}{\includegraphics[width=0.5\columnwidth]{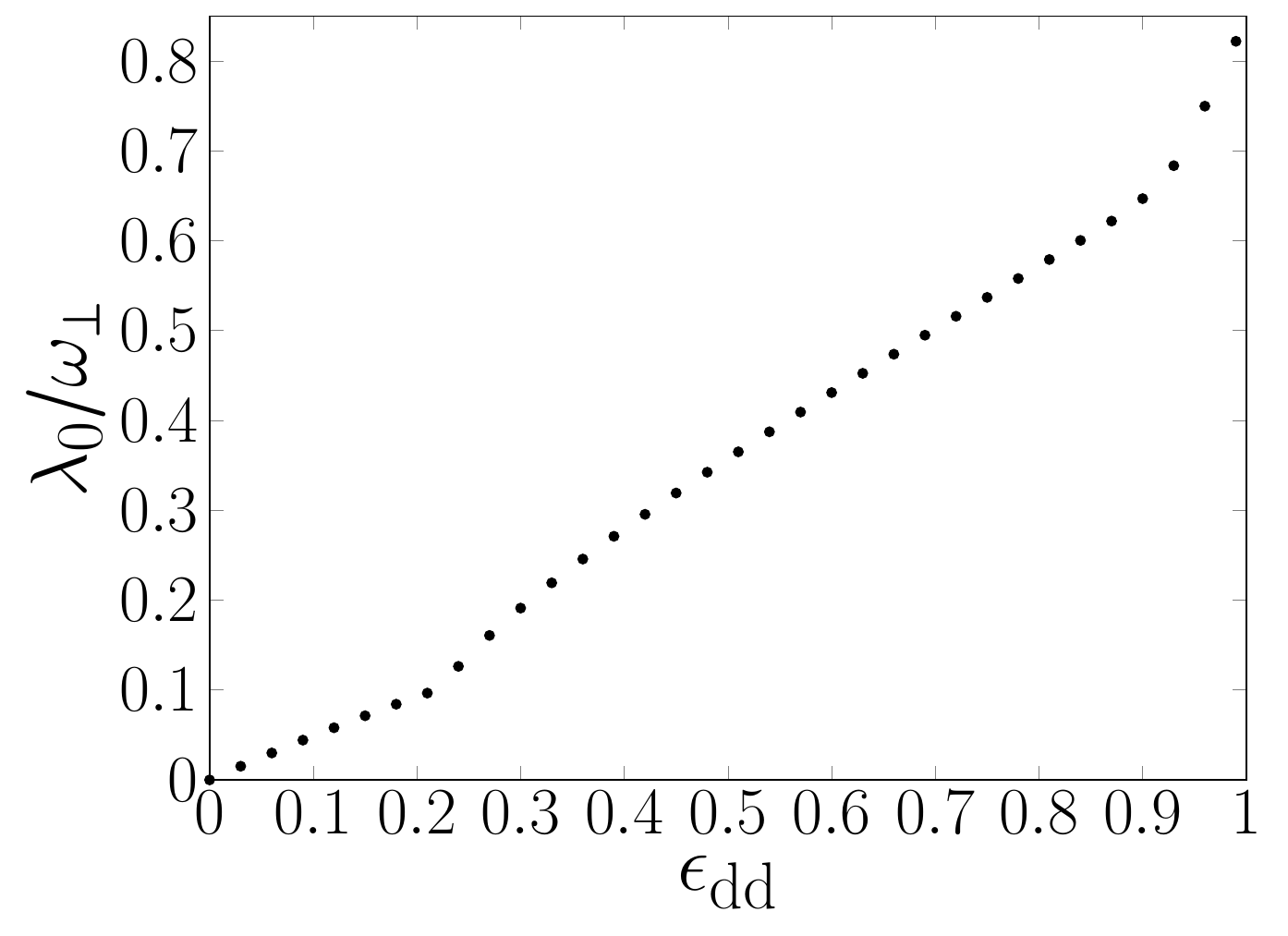}}}}
\vspace*{-5mm}
\caption{Dynamical instability of the TF stationary states: $\sqrt[4]{\lambda_0}$ as a function of $\Omega$ and $\epsilon_{\text{dd}}$ with $\gamma = 1$ and $N_{\text{max}} = 13$. The prevalence of real, positive eigenvalues for $\epsilon_{\text{dd}} > 0$ indicates the existence of a dynamical instability. Note that black corresponds to $\sqrt[4]{\lambda_0}=0$. The inset takes a cross-section at $\Omega = 3\omega_{\perp}$, corresponding to Figs.~\ref{simulationimages} and \ref{gpevstf}, and plots $\lambda_0$, showing that $\lambda_0 > 0\,\forall\, \epsilon_{\text{dd}} > 0$.}
\label{maxeigNmax14}
\end{figure}

We limit our analysis to the regions of parameter space explored in Figs.~\ref{bothrangeomega0to6profile} and \ref{gpevstf}. As several modes might be unstable at a given point in parameter space, we merely work with the eigenvalue with the largest positive, real component, denoted by $\lambda_0$. Figure \ref{maxeigNmax14} plots $\lambda_0^{1/4}$ as a function of both $\epsilon_{\text{dd}}$ and $\Omega$, with the inset demonstrating how $\lambda_0$ behaves for the parameter space explored in the dGPE simulation via a cross-section at fixed $\Omega = 3\omega_{\perp}$.

As $\epsilon_{\text{dd}}\rightarrow 1$, more regions of parameter space become dynamically unstable to collective modes of polynomial order less than 14. In addition, the region that remains stable against these modes becomes smaller for larger $\Omega$. The inset shows that for all nonzero $\epsilon_{\text{dd}}$ (with $\Omega = 3\omega_{\perp}$ and $\gamma = 1$) the corresponding TF stationary solutions suffer dynamical instabilities, with the corresponding values of $\lambda_0$ growing as $\epsilon_{\text{dd}}\rightarrow 1$. We also note that the amplitude of a mode corresponding to a real, positive eigenvalue $\lambda_0$, increases with an associated exponential timescale of $\Omega/(2\pi\lambda_0)$ rotation cycles. For $\epsilon_{\text{dd}} = 0.1$, we find that $\lambda_0\approx 0.05$, suggesting that the exponential timescale for amplitude growth is approximately $10$ rotation cycles. For current experimental scenarios involving strongly dipolar species such as $^{164}$Dy, a Feshbach resonance is generally utilised to tune $\epsilon_{\text{dd}}$ to be slightly lower than $1$~\cite{prl_107_19_190401_2011, nature_530_7589_194-197_2016}. Figure~\ref{maxeigNmax14} shows that for $\Omega/\omega_{\perp}=6$ the dynamical instability manifests itself over a timescale of $1/\omega_{\perp}$ in a cylindrically symmetric trap. Decreasing $\Omega$ increases the timescale for manifestation of the instability.

Our method does not allow for direct modelling of the experimental report of rotationally-tuned dipolar BECs by Tang et al.~\cite{prl_120_23_230401_2018}, since the relevant trap is anisotropic in the laboratory frame. This would preclude the existence of stationary states in the rotating frame, which form the basis of our analysis. However, our formalism does describe a dipolar BEC in a cylindrically symmetric harmonic trap, with the radial trapping frequency matching the average $x-y$ trapping frequency of that experiment. Therefore, we expect that accounting for the trapping ellipticity in the $x-y$ plane would amount to a correction to the results obtained via our formalism. In this approximation, the timescale of the experiment is of the order of the typical timescale that we predict for the onset of the instability which, consequently, may not have fully manifested itself during the experimental observations. Nevertheless, the enhanced dissipation observed in the experiment may be linked to the onset of this instability and thus warrants further study. We also note that the considerable deviation from the theoretically predicted TF aspect ratio in the time-of-flight measurement for $\varphi = \pi/2$, the angle explored in our study, may be due to the presence of the instability.

In this work, we have examined the rotational tuning of harmonically-trapped dipolar BECs, explicitly accounting for the rotation of the dipole polarization. This is performed by a semi-analytical formalism for obtaining the TF stationary states of a dipolar BEC, validated through dGPE simulations. In the high-rotation frequency limit, the solutions converge to those of a dipolar BEC with a static time-averaged DDI potential, as previously predicted. Crucially, however, these solutions are dynamically unstable, with collective oscillations growing exponentially from perturbations. This prevents the formation of a stable long-lived rotationally-tuned BEC. Our findings are of importance to the large body of theoretical work which has considered the rotationally-tuned regime by assuming the robustness of the gas with the time-averaged DDI. They also suggest that experiments seeking to realize rotational tuning states must be carefully designed to avoid the seeding of instabilities and further dGPE studies should be carried out where the TF regime is not applicable. Finally, the TF formalism presented in this work may be generalized to account for an alignment of the dipoles at an arbitrary angle, $\varphi$, to the rotation axis. 

\textit{Acknowledgments} --- S.B.P. is supported by an Australian Government Research Training Program Scholarship and by the University of Melbourne. A.M.M. would like to thank the Institute of Advanced Study (Durham University, U.K.) for hosting him during the initial stages of developing this collaborative research project and the Australian Research Council (Grant No. LE180100142) for support. T.B. and N.G.P. thank the Engineering and Physical Sciences Research Council of the UK (Grant No. EP/M005127/1) for support.

\setcounter{equation}{0}
\setcounter{figure}{0}
\setcounter{table}{0}
\renewcommand{\theequation}{A\arabic{equation}}
\renewcommand{\thefigure}{A\arabic{figure}}

\begin{center}
\small{\uppercase{\textbf{Appendix A: Derivation of the Thomas-Fermi Stationary Solutions}}}
\end{center}
For a dipolar BEC of a single species of mass $m$, the corresponding dipolar Gross-Pitaevskii equation (dGPE) is given, in a reference frame rotating about the laboratory frame with an angular velocity $\Omega\hat{z}$, by
\begin{equation}
i\hbar\frac{\partial\psi}{\partial t} = \left[-\frac{\hbar^2\nabla^2}{2m} + V_{\text{T}} + V_{\text{int}} + i\hbar\Omega\left(x\frac{\partial}{\partial y}-y\frac{\partial}{\partial x}\right)\right]\psi, \label{eq:gpeap}
\end{equation}
where
\begin{gather}
V_{\text{int}}(\mathbf{r},t) = g\left|\psi(\mathbf{r},t)\right|^2 + \int\mathrm{d}\mathbf{r'}\,U_{\text{dd}}(\mathbf{r}-\mathbf{r}')\left|\psi(\mathbf{r'},t)\right|^2, \label{eq:ddpotap} \\
U_{\text{dd}}\left(\mathbf{r}\right) = \frac{C_{\text{dd}}}{4\pi}\frac{1-3\left(\hat{\mathbf{e}}\cdot\mathbf{r}\right)^2}{|\mathbf{r}|^3}. \label{eq:ddintap}
\end{gather}
In the main paper, the definitions $g = 4\pi\hbar^2a_{\text{s}}/m$, where $a_{\text{s}}$ is the atomic $s$-wave scattering length, and $C_{\text{dd}} = \mu_0\mu_{\text{d}}^2$, where $\mu_{\text{d}}$ is the bosonic dipole moment, are given. Throughout the paper, a trapping potential of the form
\begin{equation}
V_{\text{T}}(\mathbf{r}) = \frac{m\omega_{\perp}^2}{2}(x^2 + y^2 + \gamma^2z^2) \label{eq:vtrapap}
\end{equation}
is considered.

We work in the hydrodynamic formalism, obtained by recasting $\psi$ in terms of the  \textit{number density}, $n$, and \textit{phase}, $S$, of the condensate:
\begin{equation}
\psi\left(\mathbf{r},t\right) = \sqrt{n\left(\mathbf{r},t\right)}\exp\left[iS\left(\mathbf{r},t\right)\right], \label{eq:densityphaseap}
\end{equation}
The density $n$ is normalized to the condensate number, $N$, via $\int\mathrm{d}^3r\,n(\mathbf{r}) = N$. Similarly, stationary solutions of the dGPE obey
\begin{equation}
\psi\left(\mathbf{r},t\right) =\psi\left(\mathbf{r},t = 0\right)\exp(-i\mu t/\hbar), \label{eq:statdefap}
\end{equation}
where $\mu$ represents the condensate's chemical potential. A consequence of this reformulation is the identification of the laboratory-frame velocity field, $\mathbf{v}$, with
\begin{equation}
\mathbf{v} = \frac{\hbar\nabla S}{m}. \label{eq:velocityphaseap}
\end{equation}

By substituting Eqs. \eqref{eq:densityphaseap} \& \eqref{eq:velocityphaseap} into the dGPE, we obtain the dipolar superfluid hydrodynamic equations:
\begin{align}
m\frac{\partial\mathbf{v}}{\partial t} = &-\nabla\left[\frac{1}{2}m\mathbf{v}^2 + V_{\text{T}} + V_{\text{int}} - m\mathbf{v}\cdot\left(\mathbf{\Omega}\times\mathbf{r}\right)\right] \nonumber \\
 &+\left(\nabla\frac{\hbar^2}{2m\sqrt{n}}\nabla^2\sqrt{n}\right), \label{eq:eulerap} \\
\frac{\partial n}{\partial t} = &\nabla\cdot\left[n\left(\mathbf{v}-\mathbf{\Omega}\times\mathbf{r}\right)\right]. \label{eq:continuityap}
\end{align}
In the \textit{Thomas-Fermi} (TF) limit \cite{pra_78_4_041601r_2008}, where the zero-point kinetic energy of the condensate is negligible, we ignore the `quantum pressure term' proportional to $\nabla^2(\sqrt{n})/\sqrt{n}$ in Eq.~\eqref{eq:eulerap}, resulting in an Euler-like equation,
\begin{equation}
m\frac{\partial\mathbf{v}}{\partial t} = -\nabla\left[\frac{1}{2}m\mathbf{v}^2 + V_{\text{T}} + V_{\text{int}} - m\mathbf{v}\cdot\left(\mathbf{\Omega}\times\mathbf{r}\right)\right]. \label{eq:tfeulerap}
\end{equation}

Initially, we seek to find stationary solutions to Eq.~\eqref{eq:continuityap} and \eqref{eq:tfeulerap}, i.e. $\partial_t\mathbf{v} = \partial_tn = 0$. Combining this condition with Eq. \eqref{eq:statdefap} and \eqref{eq:velocityphaseap} yields
\begin{gather}
\mu = \frac{m}{2}\left[\mathbf{v}^2 + \omega_{\perp}^2\left(x^2 + y^2 + \gamma^2 z^2\right)\right] + V_{\text{int}} - m\mathbf{v}\cdot\left(\mathbf{\Omega}\times\mathbf{r}\right), \label{eq:eulerstatap} \\
\nabla\cdot\left[n\left(\mathbf{v}-\mathbf{\Omega}\times\mathbf{r}\right)\right] = 0. \label{eq:contstatap}
\end{gather}
In the TF limit, the density, $n_{\text{TF}}$, is of the form
\begin{equation}
n_{\text{TF}}\left(\mathbf{r}\right) = n_0\left(1-\frac{x^2}{\kappa_x^2R_z^2}-\frac{y^2}{\kappa_y^2R_z^2}-\frac{z^2}{R_z^2}\right), \label{eq:tfdensityap}
\end{equation}
where
\begin{equation}
n_0 = \frac{15N}{8\pi\kappa_x\kappa_yR_z^3}, \label{eq:tfnormap}
\end{equation}
is the peak density, occurring at $\mathbf{r} = 0$. For the phase, $S_{\text{TF}}$, a quadrupolar ansatz of the form~\cite{prl_86_3_377-390_2001, prl_87_19_190402_2001}
\begin{equation}
S_{\text{TF}} = (m\alpha xy-\mu t)/\hbar, \label{eq:tfphaseap}
\end{equation}
corresponding to a velocity
\begin{equation}
\mathbf{v} = \alpha\nabla(xy) = \alpha(y\hat{x} + x\hat{y}), \label{eq:velocityap}
\end{equation}
is appropriate. Substituting Eq. \eqref{eq:tfdensityap} and \eqref{eq:velocityap} into \eqref{eq:contstatap} yields a stationary solution condition for the velocity field amplitude, $\alpha$, in terms of $\kappa_x$ and $\kappa_y$:
\begin{equation}
\frac{\alpha + \Omega}{\kappa_x^2} + \frac{\alpha - \Omega}{\kappa_y^2} = 0 \Rightarrow \alpha = \frac{\kappa_x^2 - \kappa_y^2}{\kappa_x^2 + \kappa_y^2}\Omega. \label{eq:alphadefnap}
\end{equation}

Recasting $V_{\text{int}}$ in the equivalent form
\begin{gather}
V_{\text{int}}\left(\mathbf{r},t\right) = g(1-\epsilon_{\text{dd}})n\left(\mathbf{r},t\right)-3g\epsilon_{\text{dd}}\left(\hat{\mathbf{e}}\cdot\nabla\right)^2\phi\left(\mathbf{r},t\right), \label{eq:ddpotfinalap} \\
\phi\left(\mathbf{r},t\right) = \frac{1}{4\pi}\int\mathrm{d}^3r'\,\frac{n\left(\mathbf{r}',t\right)}{\left|\mathbf{r}-\mathbf{r}'\right|}, \label{eq:pseudopotap}
\end{gather}
where $\hat{\mathbf{e}} = \hat{x}$, allows for an elegant solution for the self-interaction potential corresponding to a TF density profile, for which $\phi\left(\mathbf{r}\right)$ has been found to be given by \cite{prl_92_25_250401_2004, pra_71_3_033618_2005, pra_80_3_033617_2009}
\begin{widetext}
\begin{equation}
\phi(x,y,z) =\frac{n_0\kappa_x\kappa_y}{4}\left\lbrace\frac{\beta_{000}}{2} - x^2\beta_{100}-y^2\beta_{010}-z^2\beta_{001} + \frac{x^4\beta_{200} + y^4\beta_{020}+z^4\beta_{002} + 2\left(x^2y^2\beta_{110}+y^2z^2\beta_{011}+x^2z^2\beta_{101}\right)}{2R_z^2}\right\rbrace, \label{eq:pseudopotbetaap}
\end{equation}
\end{widetext}
where $\beta_{ijk}$ are functions of $\kappa_x$ and $\kappa_y$, and are specified via
\begin{equation}
\beta_{ijk}\left(\kappa_x,\kappa_y\right) = \int_0^{\infty}\frac{\mathrm{d}\chi}{(\kappa_x^2+\chi)^{i+\frac{1}{2}}(\kappa_y^2+\chi)^{j+\frac{1}{2}}(1+\chi)^{k+\frac{1}{2}}}. \label{eq:betaap}
\end{equation}
The solution of Eq. \eqref{eq:eulerstatap}, via Eq. \eqref{eq:tfdensityap}, \eqref{eq:velocityap}, \eqref{eq:ddpotfinalap} and \eqref{eq:pseudopotbetaap}, yields the stationary state Thomas-Fermi density,
\begin{widetext}
\begin{equation}
n_{\text{TF}}\left(\mathbf{r}\right) = \frac{\mu - \frac{m}{2}\left(\widetilde{\omega}_x^2x^2+\widetilde{\omega}_y^2y^2+\gamma^2\omega_{\perp}^2z^2\right)}{g(1-\epsilon_{\text{dd}})} + \frac{3\epsilon_{\text{dd}}n_0\kappa_x\kappa_y\left(3\beta_{200}x^2+\beta_{110}y^2+\beta_{101}z^2-\beta_{100}R_z^2\right)}{2(1-\epsilon_{\text{dd}})R_z^2}, \label{eq:tffullap}
\end{equation}
\end{widetext}
with dressed trapping frequencies, $\widetilde{\omega}_x$ and $\widetilde{\omega}_y$, given by
\begin{align}
\widetilde{\omega}_x^2 &= \omega_{\perp}^2 + \alpha^2 - 2\alpha\Omega, \label{eq:omegaxap} \\
\widetilde{\omega}_y^2 &= \omega_{\perp}^2 + \alpha^2 + 2\alpha\Omega. \label{eq:omegayap}
\end{align}
By comparing the coefficients of $x^2, y^2, z^2$ in Eq. \eqref{eq:tfdensityap} and \eqref{eq:tffullap}, we find that
\begin{gather}
\kappa_x^2 = \frac{1}{\zeta}\left(\frac{\omega_{\perp}\gamma}{\widetilde{\omega}_x}\right)^2\left[1 + \epsilon_{\text{dd}}\left(\frac{9}{2}\kappa_x^3\kappa_y\beta_{200}-1\right)\right], \label{eq:kappaxap} \\
\kappa_y^2 = \frac{1}{\zeta}\left(\frac{\omega_{\perp}\gamma}{\widetilde{\omega}_y}\right)^2\left[1 + \epsilon_{\text{dd}}\left(\frac{3}{2}\kappa_y^3\kappa_x\beta_{110}-1\right)\right], \label{eq:kappayap}  \\
R_z^2 = \frac{2gn_0}{m\gamma^2\omega_{\perp}^2}\zeta, \label{eq:rz2ap} \\
\zeta = 1 + \epsilon_{\text{dd}}\left(\frac{3}{2}\kappa_x\kappa_y\beta_{101} - 1\right). \label{eq:zetaap}
\end{gather}
Substituting Eq. \eqref{eq:kappaxap} and \eqref{eq:kappayap} into Eq. \eqref{eq:alphadefnap} also yields
\begin{align}
0 &= (\alpha + \Omega)\left[\widetilde{\omega}_x^2 - \frac{9}{2}\epsilon_{\text{dd}}\frac{\omega_{\perp}^2\kappa_x\kappa_y\gamma^2}{\zeta}\beta_{200}\right] \nonumber \\
&+ (\alpha - \Omega)\left[\widetilde{\omega}_y^2 - \frac{3}{2}\epsilon_{\text{dd}}\frac{\omega_{\perp}^2\kappa_x\kappa_y\gamma^2}{\zeta}\beta_{110}\right]. \label{eq:alphakappaap}
\end{align}
Equations \eqref{eq:kappaxap}, \eqref{eq:kappayap} and \eqref{eq:alphakappaap}, together with Eq. \eqref{eq:zetaap}, fully specify the shape of the stationary TF density, via $\kappa_x$ and $\kappa_y$, and the corresponding laboratory-frame velocity field, via $\alpha$. It is clear that the parameters $g$ and $N$ enter the dimensions of the condensate through only the TF radius $R_z$, and do not affect the velocity field, $\mathbf{v}$, at all. These equations are subsequently solved self-consistently to yield $\alpha$, $\kappa_x$, and $\kappa_y$ for a given choice of $\Omega/\omega_{\perp}$, $\gamma$ and $\epsilon_{\text{dd}}$.

\vspace{5pt}
\begin{center}
\small{\uppercase{\textbf{Appendix B: Derivation of the Operator Defining the Propagation of Linear Perturbations}}}
\end{center}
\vspace{5pt}
We write the time-dependent density and phase as fluctuations about the respective stationary state values:
\begin{align}
n\left(\mathbf{r},t\right) &= n_{\text{TF}}\left(\mathbf{r}\right) + \delta n\left(\mathbf{r},t\right); \label{eq:nperturbap} \\
S\left(\mathbf{r},t\right) &= S_{\text{TF}}\left(\mathbf{r},t\right) + \delta S\left(\mathbf{r},t\right). \label{eq:sperturbap}
\end{align}
Subsequently, Eq. \eqref{eq:nperturbap} and \eqref{eq:sperturbap} are substituted into Eq.~\eqref{eq:continuityap} and \eqref{eq:tfeulerap}, which are linearized by discarding all terms which are higher than linear order in $\delta n $ and $\delta S$. This results in a system of coupled, first-order equations, given as
\begin{gather}
\frac{\partial}{\partial t}
\begin{pmatrix}
\delta S \\
\delta n
\end{pmatrix}
=
\mathcal{L}
\begin{pmatrix}
\delta S \\
\delta n
\end{pmatrix}, \label{eq:perteqnsap}
\end{gather}
where
\begin{gather}
\mathcal{L} = -
\begin{pmatrix}
\mathbf{v}_c\cdot\nabla & \frac{g}{\hbar}\left(1+\epsilon_{\text{dd}}\widehat{K}\right) \\
\frac{\hbar}{m}\nabla\cdot\left(n_{\text{TF}}\nabla\right) & \mathbf{v}_c\cdot\nabla
\end{pmatrix}, \label{eq:pertmatrixap}
\end{gather}
with
\begin{equation}
\mathbf{v}_c = \mathbf{v} - \mathbf{\Omega}\times\mathbf{r} \label{eq:equibveffap}
\end{equation}
and
\begin{equation}
\widehat{K}\left[\delta n\right] = \frac{-3}{4\pi}\frac{\partial^2}{\partial x^2}\int_{\Gamma}\mathrm{d}^3s\,\frac{\delta n\left(\mathbf{s},t\right)}{\left|\mathbf{r}-\mathbf{s}\right|} - \delta n. \label{eq:koperatorap}
\end{equation}
Note that in Eq. \eqref{eq:koperatorap}, $\Gamma$ is defined only as the domain in which the unperturbed density, $n_{\text{TF}}$, is positive, i.e. the ellipsoid whose semi-axes are given by $R_x$, $R_y$, and $R_z$. We do not extend the domain $\Gamma$ to $\Gamma = \mathbb{R}^3\,\backslash\,\lbrace n_{\text{TF}} + \delta n < 0\rbrace$, as the domain extension itself represents an effect proportional to $\delta n$, and so integrating the function $\delta n/|\mathbf{r}-\mathbf{r}'|$ over the domain extension would amount to considering $O(\delta n^2)$ effects~\cite{pra_82_3_033612_2010, pra_82_5_053620_2010}.

In the main paper, we specify a basis for $\delta S$ and $\delta n$, of the form $\lbrace x^iy^jz^k\rbrace\,:\,i + j + k < N_{\text{max}}$, for diagonalizing $\mathcal{L}$. An extensive discussion on computing the integrals that arise from substituting elements of this basis for $\delta n$ in Eq.~\eqref{eq:koperatorap} is given in Appendix B of van Bijnen, \text{et al.}~\cite{pra_82_3_033612_2010}. To determine the spectrum for a given choice of $\Omega/{\omega_{\perp}}$, $\gamma$ and $\epsilon_{\text{dd}}$, we first solve self-consistently for the stationary solutions as specified by Eqns. \eqref{eq:kappaxap}, \eqref{eq:kappayap} and \eqref{eq:alphakappaap}. These stationary solutions are then used to determine $\mathcal{L}$ and diagonalization is subsequently carried out to yield the corresponding spectrum.

\vspace{5pt}
\begin{center}
\small{\uppercase{\textbf{Appendix C: Details of Numerical Set-up}}}
\end{center}
\vspace{5pt}
We numerically solve the dGPE on a 192$^3$ grid with spatial step $d = 0.15\sqrt{\hbar/(m\omega_{\perp})} \equiv 0.15l_{\perp}$ and temporal step {$\Delta t = 0.004\omega_\perp^{-1}$}. We use the ADI-TSSP method \cite{bao_wang_2006}, which is an extension of the common split-step Fourier method to incorporate rotation. In order to reduce the effects of alias copies induced by fast Fourier transform algorithms we employ a spherical cut-off to the dipolar potential, Eq.~\eqref{eq:ddintap}, restricting the range of the DDI to a sphere of radius $R_c$ as 
\begin{align}
U_\text{dd}^{R_c}(\textbf{r})=
\begin{cases}
\frac{C_\text{dd}}{4\pi}\frac{1-3\left(\hat{\mathbf{e}}\cdot\mathbf{r}\right)^2}{r^3}\,,& r<R_c\,, \\
0\,, & \text{otherwise}.
\end{cases} \label{eq:ddicutoffap}
\end{align}
As long as we choose $R_c>L$, where $L$ is the system size, this potential is physically reasonable. The analytical Fourier transform is \cite{ronen_bortolotti_2006}
\begin{align}
\tilde{U}_\text{dd}^{R_c}(\textbf{k}) = &\frac{C_\text{dd}}{3}\left[1+3\frac{\cos\left(R_ck\right)}{R_c^2k^2}-3\frac{\sin\left(R_ck\right)}{R_c^3k^3}\right] \nonumber \\
&\times\left(3\cos^2\theta_k-1\right), \label{eq:ddikspacecutoffap}
\end{align}
where $\theta_k$ is the angle between $\textbf{k}$ and the direction of the dipoles.

In order to simulate one run we employ the following procedure. At $t = 0$, the stationary state for $\epsilon_\text{dd} = 0$, and fixed $\lbrace\Omega/\omega_{\perp}, \gamma, N\rbrace$, is obtained by evolving the dipolar GPE in imaginary time~\cite{kinetrelatmod_6_1_1-135_2013}. For $t > 0$, the real-time evolution of the dipolar GPE is accompanied by an increase of $\epsilon_\text{dd}$ increased at a rate $\mathrm{d}\epsilon_\text{dd}/\mathrm{d}t = 10^{-3}\omega_{\perp}$, allowing the realisation of stationary solutions at finite values of $\epsilon_\text{dd}$. To model random external symmetry-breaking perturbations, which may shift the condensate state away from the stationary state in an experimental scenario, the condensate density is modified at the initial timestep with the addition of a random, local perturbation of up to $5\%$ of the density at each spatial grid point.

%

\end{document}